# The Integration of Agile Methodologies in DevOps Practices within the Information Technology Industry

Ridewaan Hanslo Dr

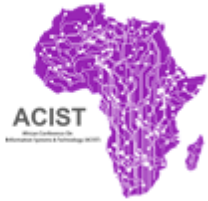

# The Integration of Agile Methodologies in DevOps Practices within the Information Technology Industry

**Ashley Hourigan**
University of Pretoria
hourigan89@gmail.com

**Ridewaan Hanslo**
University of Johannesburg
ridewaanh@uj.ac.za

## ABSTRACT

The demand for rapid software delivery in the Information Technology (IT) industry has significantly intensified, emphasising the need for faster software products and service releases with enhanced features to meet customer expectations. Agile methodologies are replacing traditional approaches such as Waterfall, where flexibility, iterative development and adaptation to change are favoured over rigid planning and execution. DevOps, a subsequent evolution from Agile, emphasises collaborative efforts in development and operations teams, focusing on continuous integration and deployment to deliver resilient and high-quality software products and services. This study aims to critically assess both Agile and DevOps practices in the IT industry to identify the feasibility and applicability of Agile methods in DevOps practices. Eleven semi-structured interviews were conducted with Agile and DevOps practitioners in varying capacities across several sectors within the IT industry. Through thematic analysis, 51 unique codes were extracted and synthesised into 19 themes that reported on each phase of the DevOps lifecycle, specifically regarding the integration and implementation of Agile methods into DevOps practices. Based on the findings, a new understanding detailing the interrelationship of Agile methods in DevOps practices was discussed that met the research objectives.

## INTRODUCTION

### Background

Within the IT industry, the demand for software products to reach consumers and markets faster has grown significantly over time (Banica et al., 2017). Consumer and market expectations are that these products will be available in shorter periods with new and improved features, functionality and benefits that are in line with changing and evolving market conditions (Banica et al., 2017; Fitzgerald and Stol, 2015). Those who can deliver software products and services to market faster to meet consumer demand position themselves to gain a competitive advantage over their competitors (Erich et al., 2017). The notion of "release early, release often" is a well-established and considered concept in today's software development and delivery processes (Fitzgerald and Stol, 2015). A key benefit of this concept in practice is that it offers increased quality of products, the ability to meet customer demands with more relevance, and the ability to achieve higher quality and consistency of outputs (Fitzgerald and Stol, 2015; Sreenivasan and Kothandaraman, 2017). Nurdiana et al. (2019) mention that the project management methodologies used to support and coordinate software product development and delivery efforts within the IT Industry have evolved. With this evolution, traditional methods such as Waterfall are being replaced by more popular Agile methods that favour flexibility rather than rigid planning, requirements gathering and delivery practices (Nurdiana, et al. 2019). Further steps have been taken in this discipline with the introduction of Development Operations (DevOps), which addresses the need to validate and deliver software projects faster to live environments (Banica et al. 2017). From the departure of traditional plan-based methods, Agile represents a significant shift in discipline. It constitutes a set of practices for software development and delivery rather than plan-based approaches (Dyba and Dingsøyr, 2008). DevOps was developed in response to the lack of collaboration around software development and deployment, as well as operations teams, that Agile did not provide (Lwakatare et al., 2019; Joseph, 2017; Fitzgerald and Stol, 2015). The organisational shift that DevOps provides focuses on cross-functional teamwork surrounding continuous feature delivery, rather than distributed and siloed groups that perform functions separately (Ebert et al., 2016).

**Problem Statement**

The development and delivery of software products have seen a substantial change in recent years, according to Nurdiana et al. (2019). Most notably, to get a minimum viable product (MVP) to market faster, large-scale programmes have been broken down into smaller, more manageable software release cycles or iterations (Banica et al., 2017; Fitzgerald and Stol, 2015). To achieve a faster time to market, there is a growing demand for enhanced communication within software development teams and between software development and operations teams (Lwakatare et al., 2019; Joseph, 2017; Fitzgerald and Stol, 2015). According to the literature, DevOps stemmed from Agile methods (Leite et al., 2019) to address the shortcomings of deployment practices, as well as to address any hindrances in improvement activities (Banica et al., 2017; Jha and Khan, 2018). Gall and Pigni (2022) note that, unlike Agile, where there is common understanding and interpretation, DevOps lacks a clear conceptualisation, often leaving practitioners without a framework to guide and drive adoption. Guerrero et al. (2023) mention that while there have been efforts to seek resolution concerning terminology and conceptual differences to establish a unified DevOps framework, this often proves difficult due to a lack of consensus and consistency in concepts, relationships and definitions used by researchers and industry. Almeida et al. (2022) mention that literature exists on the benefits, challenges and alignment efforts between development and operations teams, but note that a research gap exists around the simultaneous adoption of Agile and DevOps practices, more specifically around a combined approach of Agile and DevOps. Masud et al. (2022) also suggest that pure DevOps practices cannot focus on fulfilling continuously changing user requirements by incorporating feedback in incremental cycles. Therefore, the need for Agile in DevOps becomes apparent for two overarching reasons. Firstly, it acknowledges development challenges and the need for agility. Secondly, it considers a holistic operational perspective with smaller, more manageable releases and near-immediate feedback.

## Research Question and Objective

The research question for this study is: How can DevOps teams within the IT industry integrate Agile methodologies with DevOps practices? The primary objective is to develop a new understanding of the interrelationship between Agile methods and DevOps practices, providing actionable insights for enhancing their alignment and creating a synergistic workflow.

# LITERATURE REVIEW

Project management is defined as the "application of processes, methods, knowledge, skills and experience to achieve the project objectives" (Project Management Institute, 2017). The practice has transitioned from large-scale, advanced planning to more rapid, focused and continuous delivery (Seymour and Hussein, 2014).

## The Emergence of Agile

In 2001, a working group developed and formally published the Agile Manifesto, which consists of twelve principles and four core values (Beck et al., 2001). Agile is a way of managing projects to deliver customer value through adaptive planning, rapid and frequent feedback, continuous improvement and high levels of collaboration between customers and development teams (Banica et al., 2017). In comparison to the linear approach of Waterfall, the Agile software development process is much more iterative. This allows for component development where, at each iteration, a working product is delivered that provides value to the customer.

## The Evolution to DevOps

The concept of DevOps was first introduced in 2009 by Patrick Debois (Banica et al., 2017). DevOps is a software methodology where software development (Dev) and operations (Ops) teams work together in close collaboration to deliver software products (Lwakatare et al., 2019). This was a response to the need to balance two key requirements: how to continuously release new features and how to maintain customer service without interruption (Raj and Sinha, 2020). The principles of DevOps are often summarised by the CALMS framework (Wiedemann et al., 2019):

- **Culture**. Fostering shared responsibility and breaking down silos.
- **Automation**. Automating the build, test, and deployment pipeline.
- **Lean**. Minimising waste and focusing on value-adding activities.
- **Measurement**. Collecting data on all aspects of the lifecycle to drive improvement.
- **Sharing**. Ensuring open communication and knowledge transfer between teams.

The DevOps lifecycle (Figure 1) operates as a continuous loop through planning, coding, building, testing, releasing, deploying, operating, and monitoring, with automation being a key enabler at each stage.

## Agile and DevOps: Relationship and Integration

The literature suggests Agile practices can yield benefits such as improved quality of outputs (Gill et al., 2018b). One key advantage is its capacity to support teams in evolving in a dynamic way while maintaining their emphasis on providing consumers with a high-quality and valuable product (Gheorghe et al., 2020). However, Agile practices also face challenges. Hemon et al. (2020) point out that increased cadence in development can create bottlenecks with the operations functions that release the software.

Leite et al. (2019) add that Agile places little focus on deployment-specific practices, which can cause delays.

**Figure 1**

*DevOps Lifecycle (Adapted from Yarlagadda, 2021)*

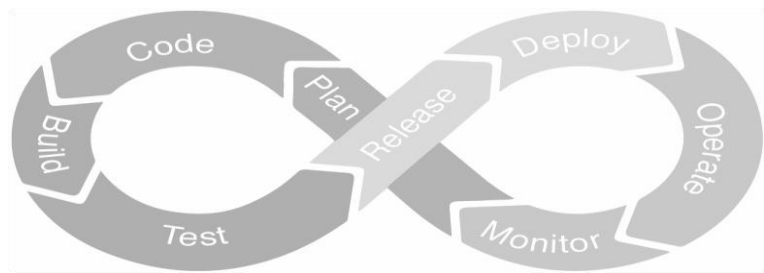

DevOps offers its benefits, including increased collaboration achieved when development and operations teams are more closely aligned (Hemon et al., 2020). Automation of integration, testing, and release also significantly reduces human error and provides greater efficiency (Banica et al., 2017; Gill, 2018a; Leite et al., 2019). Challenges in DevOps include the wide range of available technologies and tools, leading to divided opinions on its practical application (Erich et al., 2017; Hemon et al., 2020). Operations teams' worries about system stability can also lead to delays in the release of software updates (Leite et al., 2019). Agile and DevOps are compatible approaches that can cooperate to produce high-quality software (Nyale & Angolo, 2023). Nyale and Angolo (2023) add that Agile serves as a foundation for DevOps through efforts around collaboration, adaptability, and continuous improvement, and is often considered a prerequisite. Consequently, Agile influence in DevOps has evolved into a hybrid approach, incorporating methodologies such as Scrum and Kanban to facilitate planning and understand the interaction between development tasks. Methodologies such as Extreme Programming, Kanban, and Scrum are well known in Agile practice to manage workflow and fit well into the DevOps development process (Almeida et al., 2022). In conclusion, IT software development methods have changed over time, moving from conventional Waterfall methods to more iterative methods and ultimately to continuous integration and deployment (Banica et al., 2017; Hemon et al., 2020). While some authors contend that DevOps is simply Agile's progression from development to systems operation, the two approaches share a common goal of expediting application development and releasing new features frequently to produce a functional system that benefits end users (Banica et al., 2017; Gill et al., 2018a; Leite et al., 2019).

## RESEARCH METHODOLOGY

The interpretivist research philosophy was selected for this study. Interpretivism is concerned with the identification, exploration and explanation of factors in social contexts, most noticeably through subjective perspectives (Oates, 2006; Saunders et al., 2007). As this study considers human dynamics, neither positivism nor realism would be an appropriate philosophy. The interpretivist philosophy allows the researcher to capture meaning and knowledge in the differing human and social interactions within the IT industry (Ryan, 2018). An inductive research approach was chosen, which focuses on theory-building rather than testing a theory or hypothesis (Saunders et al., 2007). This approach seeks to generate meaning from collected data and identify patterns and relationships to build new theories, which are better suited to the nature and objective of this research.

## Research Strategy and Method

This research study uses a qualitative mono-method. Qualitative research is better suited as it focuses mainly on non-numerical or non-quantifiable data collection and analysis (Oates, 2006). The research strategy selected is survey research. This strategy provides the opportunity to gather meaningful data from Agile and DevOps practitioners based on first-hand experience and knowledge.

## Population and Sampling

The research population includes individuals from various sectors within the IT industry, such as finance, telecommunications, and healthcare, where Agile and DevOps are practised. A cross-sectional time horizon was chosen, allowing the researcher to collect data from individuals at a particular point in time (Melnikovas, 2018). Non-probability sampling was identified as the more appropriate sampling method. Specifically, purposive sampling was used, where participants were identified and selected through the social media and business community platform LinkedIn. Approximately 12 research participants were deemed an acceptable sample size to allow for sufficient depth of understanding.

## Data Collection and Analysis

Interviews were used as the primary instrument for data collection. The interviews were semi-structured, with most questions being open-ended, which helps obtain first-hand knowledge and capture more comprehensive responses (Salkind, 2014). The interview questions are available via https://bit.ly/AppendixAInterview. All interview responses were audio-recorded and transcribed verbatim. The qualitative data analysis method used is thematic analysis, which attempts to find patterns of meaning within the collected and generated data (Braun and Clarke, 2006). This method focuses on people's experiences, views, and opinions and is considered an exploratory process. The process involves familiarising oneself with the data, generating codes, searching for themes, reviewing themes, defining and naming them, and finally, producing a report (Braun & Clarke, 2006). The primary objective of the final report is to provide a concise, coherent, logical, and non-repetitive view of the data and the narrative it presents (Braun and Clarke, 2006).

## FINDINGS AND DISCUSSION

## Participant Demographics and Background

The study included 11 participants from South Africa, the United Kingdom, and Greece, all of whom were over 18 years of age. A significant majority of the participants reported having more than five years of experience in both DevOps software development projects and the use of Agile methodologies within those projects. Most have been involved in more than ten DevOps software development projects. The participants held various mid to senior-level roles, including software developers, software engineers, technical consultants, solutions architects, and data engineers. The roles reflected a broad spectrum of responsibilities, from focusing on specific application functionality to overseeing the entire system development lifecycle. The participants came from organisations of varying sizes, including start-ups and large global corporations, which provided a diverse perspective on team resourcing. Team sizes on their most recent DevOps projects ranged from two to ten members, with an average of five to six members per team, aligning with the ideal Agile team size. The tools participants were familiar with spanned the entire DevOps lifecycle. For the planning phase, JIRA and Confluence were popular for project management and documentation. Tools like Git, Bitbucket, and Jenkins support continuous integration. Testing was facilitated by tools such as Cucumber for automated testing. Continuous delivery and deployment were managed using Octopus and Kubernetes. Various AWS and Google Cloud Platform services were utilised throughout the CI/CD pipeline, while Kibana served as a key tool for monitoring application performance.

## Thematic Analysis of Agile Integration in the DevOps Lifecycle

The data was thematically analysed following an inductive coding approach, resulting in 19 themes across the DevOps lifecycle. Figure 2 displays a theme map outlining the relationship between the themes emerging from the qualitative data identified as part of the thematic analysis process. In addition, the code syntheses to generate the themes are available at https://bit.ly/AppendixBThemeCodes.

### Agile Methods and Techniques

The most frequently identified Agile methodologies were Kanban and Scrum. Kanban was favoured for its simplicity, focus on limiting work in progress, and visual workflow management, often facilitated by tools like Jira (Matharu et al., 2015). Participant 3 noted, "*...in our team we use Kanban, which for us works the best because it allows you to focus on one thing and complete that thing before moving on to the next task...*". Scrum was also widely implemented, though often in a hybrid manner where teams selectively adopt practices like retrospectives for continuous improvement, rather than strictly adhering to all ceremonies. The Scaled Agile Framework (SAFe) was noted in larger organisations, receiving mixed feedback; it was praised for providing structure but criticised for its complexity and potential to limit flexibility. Interviewee 11 expands on her teams' preference for the SAFe methodology. "*I would prefer working in a SAFe methodology. I think it's more organised and it just gives each one clearly defined role, especially when it comes to DevOps.*". Agile development practices are centred on delivering an MVP and incrementally adding features. Collaborative approaches like Behaviour Driven Development (BDD) and Test-Driven Development (TDD) were identified as core development objectives, fostering shared understanding and ensuring functional testing.

### DevOps Planning

Planning in an Agile-DevOps context is dynamic and feedback-driven, responding to challenges from previous work cycles. This reflects an Agile mindset of adapting to change for continuous improvement. Teams often employ short-term, iterative planning rather than long-term, rigid plans. Participant 10 reflected on the Agile mindset of being able to adapt to change. "*I'd say with planning, we take an agile approach, as in, we don't, we won't try to plan out a year's work at the beginning… So I'd say we do a lot of short-term planning rather than long-term planning. We're happy to iterate. So we plan for things to go wrong. We plan for things to be unpacked and discovered.*" A common practice is "backwards scheduling," where teams work backwards from a set release date to establish milestones. User stories are critical for defining requirements in a lightweight manner, aligning with the Agile value of working software over comprehensive documentation. Participants highlighted the importance of direct interaction with stakeholders to understand priorities and adapt to changing requirements. Agile ceremonies, particularly those from Scrum like daily stand-ups, sprint planning, and retrospectives, are widely used to structure development effort, facilitate communication, and drive continuous improvement. JIRA was the predominant tool for managing backlogs, tracking progress, and providing visibility.

**Figure 2**

*Thematic Analysis Theme Map*

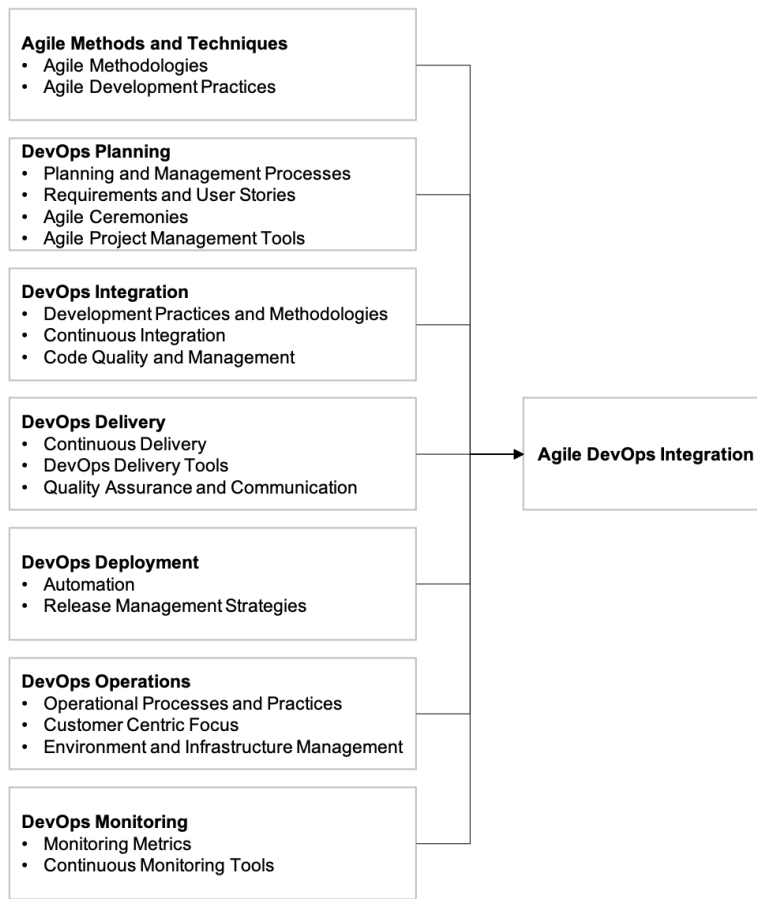

## DevOps Integration

Continuous integration (CI) is a core practice, with developers frequently committing small code changes to a shared repository (Elazhary et al., 2022). The process typically involves pull requests for peer review, followed by automated builds and tests upon merging. This iterative approach, supported by Agile frameworks like Scrum and Kanban, facilitates continuous feedback and ensures a working version of the software is always available. Interviewee 9 shows confidence in his team's integration process "*...and the multiple environments like QA and staging, basically allow us to like continuously push code without worrying about breaking much.*". Ensuring code quality is a major focus, achieved through practices like static code analysis using tools like SonarCube, and development methodologies such as BDD and TDD. Version control systems, primarily Git and Bitbucket, are essential for managing code changes and maintaining quality.

## DevOps Delivery and Deployment

Continuous delivery (CD) is achieved through a high degree of automation in the build, test, and deployment processes. While automation is favoured, some manual steps, particularly for validation, remain. The delivery process includes rigorous automated and regression testing to ensure the stability and reliability of releases. Artefacts are commonly packaged as container images and managed in repositories like AWS ECR, allowing for version tracking and easy rollbacks. Deployment automation is often managed through feature flagging, which decouples deployment from release, allowing teams to deploy latent code and activate features for specific users or environments. Research participant 10 spoke about the minimum automated effort in the team's deployment processes. "*On each merge, it usually triggers a Git pipeline that does some automated testing at the very a build test to make sure the code is building.*" This strategy provides flexibility and minimises downtime. Release management strategies vary, from frequent, small deployments to larger, fully featured releases. IT service management and formal change control processes are often used, especially for production deployments, to ensure all requirements are met before authorisation.

### DevOps Operations and Monitoring

In operations, there is a shift towards building in visibility and measurement from the project's inception. Post-implementation reviews are common for evaluating the performance of deployed solutions and gathering user feedback. A strong customer-centric focus is evident, driven by the Agile principle of delivering value (Goericke, 2020). Continuous feedback loops with customers are established to inform the development cycle. Research participant 3 elaborates on the importance of customer centricity. "*…whenever we add a new feature, we add a new test case or a few test cases that test that feature, the quality of the code then improves… but it's about the customer experience. So, when we give something to the customer, then they know that it works.*". Environment and infrastructure management increasingly rely on cloud services (IaaS) and managing configurations as code to ensure speed and flexibility. Monitoring is a continuous, tightly integrated practice (Dakkak et al., 2022). Alerts from tools like PagerDuty and Datadog are often channelled into platforms like Slack and Jira to ensure rapid response and integrate issue tracking into the development workflow. This creates a proactive approach to monitoring and a culture of continuous learning and improvement.

### Challenges and Summary

The integration of Agile methodologies into DevOps presents both benefits and challenges. A key challenge is a widespread lack of deep understanding of Agile principles beyond the popular frameworks of Scrum and Kanban. This leads to inconsistent application and difficulty in moving beyond surface-level practices. Larger organisations, in particular, struggle to move away from traditional, more rigid methods. Despite these hurdles, the research confirms that the successful adoption of Agile methodologies within DevOps is organic, with teams tailoring practices to their specific needs. The core shared principles of rapid, iterative development, continuous feedback, and delivering value to the end-user are driving this integration. The findings illustrate that Agile methodologies provide the foundational workflows and collaborative culture that enable DevOps practices to succeed, streamlining development and reinforcing a commitment to quality and efficiency. To overcome the identified challenges, a greater emphasis on education, coaching, and strategic change management is essential to foster a deeper, more effective application of Agile principles within the DevOps context.

## CONCLUSION

This study sets out to understand how DevOps teams integrate Agile methodologies into their practices. The findings reveal a deep, symbiotic relationship where Agile provides the essential cultural, procedural, and philosophical framework that allows the technical automation of DevOps to succeed and deliver true value. While DevOps provides the engine for speed and reliability, Agile provides the steering mechanism, ensuring the destination is aligned with customer needs and business objectives.

The integration is not about choosing one methodology over the other but about leveraging the strengths of both to create a holistic, adaptive, and efficient system for software delivery.

## Research Contribution

This study offers an empirical account of how Agile methodologies are integrated into each phase of the DevOps lifecycle. It moves beyond high-level theoretical discussions to provide concrete examples and practices directly from industry practitioners. For practitioners, it validates the use of hybrid models and offers specific, proven techniques, such as Scrum ceremonies and using Kanban for workflow management to improve their processes. For academia, this research lays the groundwork for developing more formalised models and frameworks for Agile-DevOps integration, providing empirical data to support future theoretical work.

## Limitations and Future Research

The findings of this study are derived from a qualitative analysis of 11 practitioners. While this provides depth and rich context, the results have limited generalisability. Future research could build upon these findings with larger, quantitative studies to assess the prevalence and impact of these integration practices across a wider range of industries and organisational sizes. Furthermore, a longitudinal study tracking the evolution of these hybrid practices over time would be highly valuable. Investigating the role of organisational culture, leadership, and team maturity in fostering successful hybrid Agile-DevOps environments presents another critical avenue for future inquiry. Ultimately, the successful integration of Agile and DevOps is about creating a system that balances speed with direction, and automation with human-centric adaptation. By embedding the core Agile principles of iterative development, customer focus, and continuous feedback into the powerful, automated engine of DevOps, organisations can achieve a powerful synergy that accelerates the delivery of high-quality, customer-centric software in today's demanding market.